\documentclass{PoS}

\title{New actions for minimally doubled fermions and their counterterms}

\ShortTitle{New actions for minimally doubled fermions and their counterterms}

\author{\speaker{Stefano Capitani} \\
        Institut f\"ur Kernphysik and HIM (Helmholtz-Institut Mainz), \\
        University of Mainz, Johann-Joachim-Becher-Weg 45,
        D-55099 Mainz, Germany \\
        E-mail: \email{capitan@kph.uni-mainz.de}}

\abstract{Minimally doubled fermions provide a cheap and convenient way
          of simulating quarks which preserve chiral symmetry. It has been
          established that two actions of this kind (known as Bori\c{c}i-Creutz
          and Karsten-Wilczek) require the tuning of three counterterms
          in order to be properly renormalized. Here we construct some more
          general minimally doubled actions and investigate the properties of
          their counterterms.}

\FullConference{31st International Symposium on Lattice Field Theory
                 LATTICE 2013\\
		 July 29 - August 3, 2013\\
		 Mainz, Germany}

\begin{document}

\section{Motivation}

Minimally doubled actions provide convenient fermion lattice formulations
which preserve chiral symmetry for any finite lattice spacing $a$. Two fermion
flavors are the minimal amount allowed by the Nielsen-Ninomiya theorem
in order to maintain, together with an exact continuous chiral symmetry
(of the standard type, i.e. not Ginsparg-Wilson), other convenient properties
like locality and unitarity. Such chiral fermionic formulations can still be
kept ultralocal, like Wilson fermions, but at variance with the latter no
tuning of masses is required, since the continuous chiral symmetry protects
masses from additive renormalization.

Compared with staggered fermions, which have the same kind of $U(1)$ chiral
symmetry, minimally doubled fermions are computationally slightly more
expensive, however having 2 flavors instead of 4 they do not require
uncontrolled extrapolations to 2 physical light flavors, and so they are ideal
for $N_f=2$ simulations. One also avoids the complicated intertwining of
spin and flavor of staggered fermions. As they are much cheaper than
Ginsparg-Wilson fermions, minimally doubled fermions can be very convenient
for vector-like theories like QCD. Moreover, they might be very practical for
simulations of lattice QCD at finite temperature, where staggered fermions are
extensively used (as a glance at the corresponding contributions in these 
proceedings can attest).

Two particular realizations of minimally doubled fermions have been studied 
in deeper detail in the last few years, the Karsten-Wilczek 
\cite{Karsten:1981gd,Wilczek:1987kw} and Bori\c{c}i-Creutz 
\cite{Creutz:2007af,Borici:2007kz,Creutz:2008sr,Borici:2008ym} actions. 
Since they contain only nearest-neighbor interactions, these actions are quite
cheap and easy to simulate. Moreover, they allow the construction of conserved
axial currents which have a simple expression (again only involving
nearest-neighbor sites).

It turns out that for the massless Karsten-Wilczek (KW) and Bori\c{c}i-Creutz 
(BC) formulations three counterterms need to be added to the bare action
in order to remove the hypercubic-breaking contributions
\cite{Capitani:2009yn,Capitani:2009ty,Capitani:2010nn,Capitani:2010ht}.
A consistent renormalized theory can be achieved only for special values of the
coefficients of the counterterms, which then have to be appropriately tuned.
Once these special values are known one can carry out Monte Carlo computations
of physical observables almost as easily and inexpensively as with Wilson
fermions. 

We present here minimally doubled actions that have the correct continuum limit
(like the KW and BC fermions) but require fewer counterterms when appropriate
values of their 2 parameters $\alpha$ and $\lambda$ are chosen.
These actions can be seen as generalizations of the KW action,
where the distance $2\alpha$ (modulo $2\pi$) in momentum space between the
2 poles of the quark propagator can be varied at will, like in the actions
proposed in \cite{Creutz:2010qm,Creutz:2011hy}. 

The three possible counterterms for all actions presented here (including the
next-to-nearest-neighbor actions of Sect.~\ref{sec:next}) are the same of the
standard KW action. This happens because both poles of the quark propagator
still lie entirely on the temporal axis, and thus the temporal direction is
always selected as the special one (irrespective of the values of $\alpha$ and
$\lambda$), and moreover the spinorial structure of all these actions is also
the same. Thus, $P$ is a symmetry, and also $CT$ \cite{Bedaque:2008xs}, but
$T$ and $C$ separately are violated (unless the actions are properly
renormalized). 

In massless quenched QCD only 2 of these counterterms are needed, the fermionic
counterterm of dimension four, of the form
$id_4 (g_0) \,\overline{\psi}\,\gamma_4 D_4 \psi$, 
and the counterterm of dimension three, of the form
$id_3 (g_0)/a \,\overline{\psi}\,\gamma_4 \psi$.
In full QCD the gluonic part of the action can generate through internal
quark loops a gluonic counterterm, of the form
$d_{\rm{P}}(g_0) 
\sum_{\rho,\lambda} {\rm Tr}\, F_{\rho\lambda}(x) \, F_{\rho\lambda}(x) \,
\delta_{\rho 4}.$

The values of the coefficients of the counterterms for which one is able to 
obtain a consistent renormalized theory are functions of $\alpha$ and
$\lambda$ which could vanish in some place. We indeed find that this is 
the case, and that there are a few curves in the space spanned by $\alpha$
and $\lambda$ for which one of the counterterms vanishes. Thus, the renormalized
actions corresponding to these particular choices of the parameters require
only 2 counterterms. The ultimate goal is to find actions for which all
functions happen to become zero for the same values of $\alpha$ and $\lambda$.
In this case no counterterms at all will be required, and one will be able to
carry out consistent simulations using just the tree level actions. They will
be then much cheaper than the already convenient case of (say) KW fermions.

\section{Nearest-neighbor minimally doubled actions}

We study the class of bare fermionic actions
\begin{equation}
a^4 \sum_{x} \overline{\psi} (x) \,\Big\{ 
\frac{1}{2} \sum_{\mu} \Big[ \gamma_\mu (\nabla_\mu+\nabla^\star_\mu) 
- i a \gamma_4 \,(\lambda +\delta_{\mu 4}(\cot \alpha-\lambda)) \,
\nabla^\star_\mu \nabla_\mu \Big] + m_0  \Big\} \,\psi (x) ,
\label{eq:action}
\end{equation}
where $\nabla_\mu\,\psi (x) = (U_\mu(x) \psi(x+a\widehat{\mu})
- \psi (x))/a$.\,\footnote{For the expanded expressions of these actions,
and of those of the next-to-nearest-neighbor actions introduced in
Sect.~\ref{sec:next}, see also Ref.~\cite{Capitani:2013zta}.}
These minimally doubled actions have $\mu=4$ as a special direction (like in
the standard KW action), satisfy $\gamma_5$-hermiticity, and the interactions
are only between nearest-neighbor lattice sites. The standard KW action
corresponds to $\alpha=\pi/2$.

In momentum space the Dirac operators of these fermions read, in the free case,
\begin{equation}
  \frac{i}{a} \sum_{\mu=1}^4 \gamma_\mu
   \sin ap_\mu +\frac{i\gamma_4}{a} \,
\Big[ \lambda \sum_{k=1}^3 (1 -\cos ap_k) + \cot \alpha\,(1 -\cos ap_4) \Big] 
+ m_0 ,
\label{eq:mom-action}
\end{equation}
and they have two zeros, located at $a\bar{p}_1=(0,0,0,0)$ and
$a\bar{p}_2=(0,0,0,-2\alpha)$, which describe two fermions of opposite
chirality. For $\alpha = 0$ and $\alpha = \pi$ the actions become singular 
($\cot \alpha = \infty$), and the range of $\alpha$ can be taken as 
$0 < \alpha < \pi$.

Varying $\lambda$ does not change the location of any of the zeros of the
actions, as this parameter has only the task of decoupling the 14 other
fermions from the naive fermionic action (which corresponds to the first term
in Eq.~(\ref{eq:mom-action})), giving them a mass of order $1/a$. However,
to avoid the appearance of other doublers it must also be (at tree level)
$\lambda > (1-\cos \alpha)/(2\sin \alpha)$.

All the actions presented and considered in this article have, irrespective
of the values of $\alpha$ and $\lambda$, the correct continuum limit.
Since they are Wilson-like with hopping terms of only one unit of the lattice
spacing $a$, they are rather cheap to simulate. The computational effort will be
about a few times the one required for Wilson fermions \cite{numerical}. 

The possibility of constructing a conserved axial current, which also
has a simple form and is cheap to use in Monte Carlo simulations, 
constitutes one of the major advantages of using these formulations of 
minimally doubled fermions. Its expression for the above actions is
\begin{eqnarray}
A_\mu^{\mathrm cons} (x;\alpha,\lambda) &=& \frac{1}{2} \Big(
   \overline{\psi} (x) \, (\gamma_\mu -
      i\gamma_4 \, (\lambda+\delta_{\mu 4}(\cot \alpha-\lambda)) ) \,
   \gamma_5 \, U_\mu (x) \, 
   \psi (x+a\widehat{\mu}) 
  \label{eq:noether-axial} \\ 
 && \qquad \quad + \overline{\psi} (x+a\widehat{\mu}) \, 
    (\gamma_\mu +
      i\gamma_4 \, (\lambda+\delta_{\mu 4}(\cot \alpha-\lambda)) ) \,
  \gamma_5 \, U_\mu^\dagger (x) \, \psi (x) \Big) \nonumber \\
 && + \frac{d_4 (g_0)}{2} \Big(
   \overline{\psi} (x) \, \gamma_4 \gamma_5 \, U_4 (x) \, 
   \psi (x+a\widehat{4}) 
  + \overline{\psi} (x+a\widehat{4}) \, \gamma_4 \gamma_5 
   \, U_4^\dagger (x) \, \psi (x) \Big) . \nonumber
\end{eqnarray}
This is an exactly conserved quantity for any finite value of the lattice 
spacing $a$, and only involves nearest-neighbor sites. This is particularly
important, as not many fermionic formulations exist for which a conserved
axial current exists and is of such a simple form.

\section{Curves of counterterm removal} 

Counterterms need to be introduced to compensate (when properly tuned) for
hypercubic-breaking factors. Their coefficients can be determined by making 
these factors disappear in 1-loop amplitudes, so that these actions are
properly renormalized. The main objective of this work is to see if and when
the counterterms can vanish. This means that the corresponding functional forms
are missing in the 1-loop radiative corrections when special values of
$\alpha$ and $\lambda$ are employed. We use Wilson's plaquette action in a
general covariant gauge (where $\partial_\mu A_\mu=0$), and $m_0=0$.

Due to the non-trivial form of the denominator of the quark propagator, 
it is not possible to provide results with an analytic dependence on
$\alpha$ or $\lambda$. The search for the special values of these parameters 
which remove the hypercubic-breaking factors in the 1-loop quark self-energy 
and vacuum polarization must then be carried out numerically.
The tadpole of the self-energy however can be calculated analytically, and
in a general covariant gauge is given by\,\footnote{The quantity
$Z_0=0.1549333\ldots$ is an often-recurring lattice integral
\cite{GonzalezArroyo:1981ce,Ellis:1983af,Capitani:2002mp}.}
\begin{equation}
T = g_0^2 \, C_F \, \frac{Z_0}{2} \Big(1-\frac{1}{4}(1-\xi)\Big) 
    \,\Big( i\not{\hspace{-0.08cm}p}
  - \frac{i\gamma_4}{a}\,(3\lambda+\cot \alpha) \Big) .
\label{eq:tadpole}
\end{equation}
To carry out the calculations of the other diagrams required for 
the tuning of the counterterms we have used 
a set of computer codes written in the algebraic manipulation language FORM 
\cite{Vermaseren:2000nd,Vermaseren:2008kw}, extended to include the special
features of the actions presented here.

Our main results are summarized in Fig.~1, which shows the curves for which
the various counterterms have a vanishing coefficient. Each counterterm can be
removed, but there are no intersections between the curves of zeros, and so 
at least 2 counterterms are always required.\,\footnote{It is not necessary
at this stage to determine the zeros with high precision, as we mainly want to
show that curves of zeros exist, and see what shape they have. When a 
nonperturbatively renormalized action which needs just one (or no) counterterm
will be found, a determination with higher precision of
the relevant parameters will be then desirable.}

\begin{figure}[tl]
\begin{center}
\includegraphics[height=7.1cm]{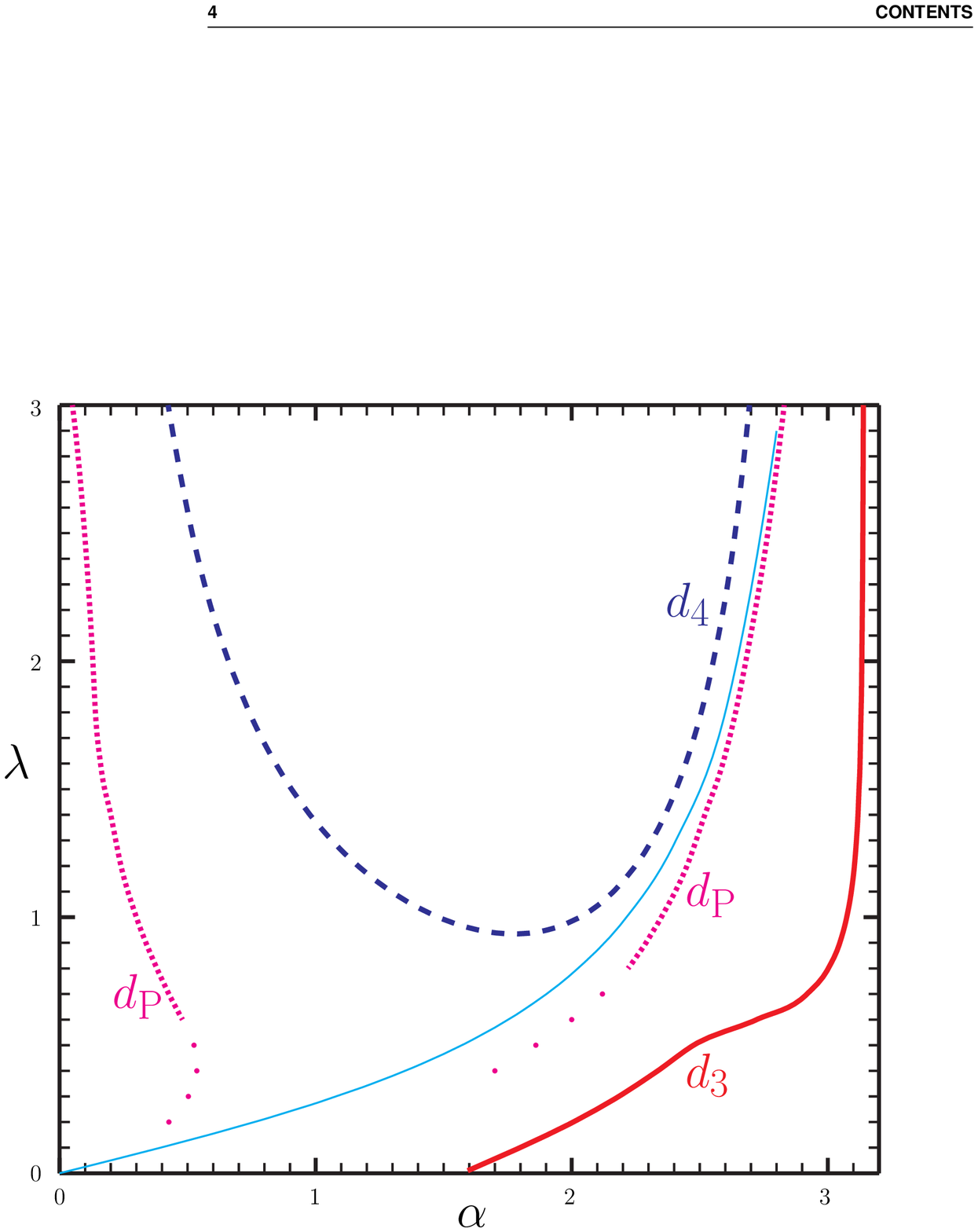}
\end{center}
\caption{The curves on which the coefficients of the various counterterms
vanish. Shown is also the function $\lambda = (1-\cos \alpha)/(2\sin \alpha)$, 
below which additional doublers can appear (at tree-level).}
\end{figure}

\section{Next-to-nearest-neighbor minimally doubled actions}
\label{sec:next}

It would be nice to find minimally doubled actions for which intersections
between the curves of zeros exist, so that 2 (or even all) of the possible
counterterms can then be removed. We introduce then interactions also between
next-to-nearest-neighbor lattice sites. It could be that actions which contain
interactions also at distance $2a$ or larger have somewhat different
properties, and the hope is that at the end some of these actions will not
require any counterterms to be properly renormalized.

We find then useful to propose here a first example of a class of minimally
doubled actions with next-to-nearest-neighbor interactions, which depend on
4 parameters:\,\footnote{Here
$\widetilde{\nabla}_\mu\,\psi (x) = (U_\mu(x) \, U_\mu (x + a\widehat{\mu})
\,\psi(x+2a\widehat{\mu}) - \psi (x))/(2a)$ is another discretization of
the lattice covariant derivative, extending this time over two lattice sites.}
\begin{equation}
a^4 \sum_{x} \overline{\psi} (x) 
\,\Big\{ \sum_{\mu} \Big[ \frac{1}{2} \,\gamma_\mu (\nabla_\mu+\nabla^\star_\mu) 
\label{eq:concisenextaction}
- i a \gamma_4 \Big\{ \frac{1}{2}\, f^{(1)}_\mu \, \nabla^\star_\mu \nabla_\mu
  -f^{(2)}_\mu \, \widetilde{\nabla}^\star_\mu \widetilde{\nabla}_\mu \Big\} \Big]
  +m_0 \Big\} \,\psi (x), 
\end{equation}
where
\begin{eqnarray}
f^{(1)}_\mu(\alpha,\lambda,\lambda',\rho) &=& \lambda+2\lambda' 
   +\delta_{\mu 4}\,\Big(\Big(\rho+ \frac{1-\rho}{\sin^2 \alpha}\Big)
   \cot \alpha-\lambda-2\lambda'\Big) , \\
f^{(2)}_\mu(\alpha,\lambda',\rho) \ \  &=& \lambda' +\delta_{\mu 4}\,\Big(
   \frac{1-\rho}{2\sin^2 \alpha} \cot \alpha-\lambda'\Big) .
\end{eqnarray}
These actions satisfy $\gamma_5$-hermiticity, and like for the standard KW
action the temporal direction is again the special one which is selected and
which then breaks hypercubic symmetry. The symmetries of these actions and
their possible counterterms are the same of the standard KW action.

The corresponding momentum-space actions are given in the free case
by\,\footnote{Notice that for $\lambda'=0$ and $\rho=1$ one falls back to the
case of the nearest-neighbor actions (\ref{eq:action}).}
\begin{eqnarray}
&& \frac{i}{a} \sum_{\mu=1}^4 \gamma_\mu
   \sin ap_\mu +\frac{i\gamma_4}{a} \, \Big\{ \sum_{k=1}^3 \Big( 
   \lambda\,(1 -\cos ap_k) + \lambda'\,(1 -\cos ap_k)^2 \Big) 
\label{eq:mom-nextaction} \\
  && \qquad \qquad \qquad \qquad \qquad \qquad
   + \cot \alpha \,\Big( \rho\,(1 -\cos ap_4) 
   +\frac{1-\rho}{2\sin^2 \alpha}\,(1 -\cos ap_4)^2 \Big) \Big\} + m_0 .
   \nonumber
\end{eqnarray}
The parameter $\alpha$ regulates the distance between the two zeros, which are 
at the same positions $a\bar{p}_1=(0,0,0,0)$ and $a\bar{p}_2=(0,0,0,-2\alpha)$
as in the nearest-neighbor actions. That there are only two zeros is certainly 
the case if $-3 \le \rho \le 1$ (as we have numerically verified), and 
$0 < \alpha < \pi$ as before. For choices of $\rho$ outside of this
range, additional zeros can in general appear, and one can still get minimally
doubled actions but only for a restricted domain of $\alpha$ (whose extension
depends on the value of $\rho$).\,\footnote{For example, with the choice
$(\alpha,\rho)=(0.1,1.1)$ and for $\vec{p}=(0,0,0)$, the action is proportional
to $\gamma_4$, and its coefficient a function of $p_4$ only, which has four
intersections with the $p_4=0$ axis.} Moreover, one must also respect the
(tree level) condition $\lambda + 2\lambda' >  -\min \, \{\sin x + \cot \alpha 
\,(\rho\,(1 -\cos x) +(1-\rho)\,(1 -\cos x)^2/(2\sin^2 \alpha)) \}/2$
to ensure that there are no more than two fermions.

The hope with these next-to-nearest-neighbor actions is that for special choices
of the parameters one could hit on renormalized actions which do not require any
counterterms. The fact that there are 4 parameters, and not just 2 as in the
nearest-neighbor actions, should result in many more curves on which the
counterterms become zero and, above all, more chances for intersections among
these curves. It could then happen that there exist some values of the
parameters for which one ends up with just one counterterm, or none at all.
Adequately exploring this larger parameter space is much more expensive 
than for the nearest-neighbor actions, and this is left for future work.

\section{Outlook}

We have shown that there are curves in the space spanned by the two parameters
$\alpha$ and $\lambda$, which define the nearest-neighbor actions that we have
introduced, on which some of the counterterms vanish, and on these curves
the counterterms needed are fewer than the 3 required for the standard massless
KW and BC actions. There are numerous choices of parameters which give just 2
counterterms, and hence only 1 in the unquenched case. Although the conclusions
presented in this work arise from perturbative calculations, it is likely that
also in numerical simulations the removal of some of the counterterms can be
accomplished for appropriate choices of the parameters. The first task will
then be to check whether the qualitative pattern of the curves that we have
found here is also reproduced nonperturbatively. 

In principle some intersection points could appear at the nonperturbative
level. If some of the curves of zeros had indeed an intersection point, the
corresponding values of $\alpha$ and $\lambda$ would provide a renormalized
minimally doubled action which requires only one counterterm. Therefore, even
though here using perturbation theory this has not occurred, it could well be
the case that nonperturbatively an intersection point does exist. This would
make possible to simulate renormalized minimally doubled actions with at most 
one counterterm.

It is also possible that even cleverer minimally doubled actions can be
constructed, which would be able to remove all possible counterterms. In this
case Monte Carlo simulations employing just the bare action will be sufficient
for the extraction of significant physical results, and no tuning of
counterterms will be required to simulate such an action. 

But even when it is not possible to remove all counterterms, it is always useful
to be able to accomplish a reduction in the dimensionality of the parameter
space of their coefficients, as this makes their numerical evaluation
easier. In particular, if there is only one counterterm left, it is much
simpler to carry out the determination of its coefficient, because one has to
deal with just a one-dimensional space instead of a multi-parameter one.

In any case, apart from the removal of counterterms, it is always useful
to have as many different minimally doubled actions as possible and keep on
trying to construct new ones, because some of them could turn out to possess
better theoretical or practical properties. The actual amount of important
quantities such as the mass difference between the $\pi^\pm$ and the $\pi^0$,
or of mass splittings within otherwise degenerate multiplets, could turn out
to be rather small for a few of these actions and not for all other ones.
In  general it can be convenient to have minimally doubled actions where the
distance between the two poles of the quark propagator can be arbitrarily
varied. Special values of this distance could also provide actions which could
turn out to be more advantageous for Monte Carlo simulations
(in that for instance they minimize some artefacts peculiar to these actions).

Thus, this work can also be considered as an inspiration to undertake further
searches for new minimally doubled actions which require a reduced number 
of counterterms, and possibly (in the best of cases) none at all.
Further theoretical investigations, and the twisted-ordering method presented
in \cite{Creutz:2010cz,Misumi:2010ea,Kimura:2011ik} (see also the overview in
\cite{Misumi:2012eh}), which can be useful for constructing other minimally
doubled actions, could suggest how to steer these searches also in new
directions. 

The next-to-nearest-neighbor actions that we have introduced could also be
taken as a starting point for a special direction in this undertaking,
especially if it turns out that next-to-nearest-neighbor interactions possses
some fundamental feature different from the nearest-neighbor case,
in particular with respect to the type of counterterms which can arise.

\end{document}